\documentclass[twoside,fleqn]{article}
\usepackage{graphicx}
\usepackage{amsmath}
\usepackage{espcrc2}

\newcommand{\beq}{\begin{eqnarray}}
\newcommand{\eeq}{\end{eqnarray}}

\newcommand{\ie}{{\it i.e.\ }}

\newcommand{\pbp}{\langle \bar{\psi} \psi \rangle}

\title{Chiral transition and deconfinement in QCD}

\author{
M. D'Elia\address[GENO]{Dipartimento di Fisica dell'Universit{\`a} di
Genova and INFN, Via Dodecaneso 33, I-16146, Genova, Italy},
A. Di Giacomo\thanks{Speaker at the Conference}\address[PISA]{Dipartimento 
di Fisica dell'Universit\`a di Pisa    and INFN, Largo Pontecorvo 2, 
I-56127 Pisa, Italy}, C. Pica\addressmark[PISA]} 

\begin{document}

\begin{abstract}
The study of QCD with two light dynamical fermions 
is of fundamental importance to understand the mechanism of
color confinement.  We present results of a numerical
investigation on the order
of the chiral phase transition with $N_f = 2$ by use of a novel
strategy in finite size scaling analysis. 
We compare the critical behaviour of the specific heat, of the
chiral susceptibility and of the equation of state 
with the possible critical behaviours. 
A second order transition
in the $O(4)$ and $O(2)$ universality classes are
excluded by our data and substantial evidence emerges for a first
order transition. Like in most of previous works we have used 
the standard staggered action with $L_t = 4$: possible
scaling violations and the need for further studies are discussed.
\end{abstract}

\maketitle

\section{Introduction}

\begin{figure}[b!]
\includegraphics*[width=\columnwidth]{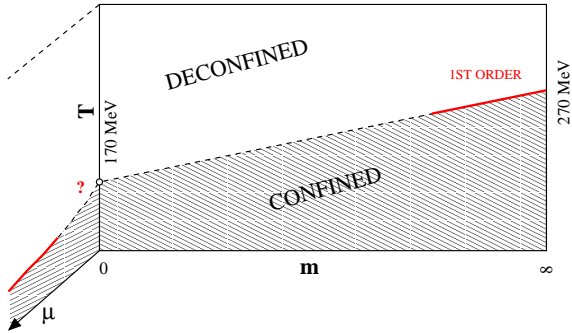}
\vspace{-30pt}
\caption{Schematic phase diagram of $N_f=2$ QCD.}\label{PHDIA}
\end{figure}

The study of the deconfinement transition in presence of two light
degenerate dynamical quarks ($N_f = 2$) is of special interest: the
system is a good approximation to Nature and it is very interesting
also from a theoretical point of view.
A sketch of the phase diagram for $N_f = 2$ 
is shown in Fig.~\ref{PHDIA}: 
$m$ is the quark mass and  $\mu$ is the baryon chemical potential.

In the $\mu = 0$ plane, 
quarks decouple in the limit $m \to \infty$ 
and the system tends to the quenched limit, where the deconfining transition
is an order-disorder first order
phase transition; $Z_3$ is an associated
symmetry and the Polyakov line
$\langle L \rangle$ is an order parameter. 
The inclusion of dynamical quarks explicitely breaks the $Z_3$ symmetry 
and $\langle L \rangle$ is not a good
order parameter, even if it works as such
for quarks masses down to $m \simeq 2.5 - 3$ GeV.

At $m \simeq 0$  a chiral phase transition takes place 
at a critical temperature $T_c\simeq 170$ MeV, from a low temperature 
phase in which  chiral symmetry is
spontaneously broken to a high temperature phase in which it is
restored: 
the corresponding order parameter in this case is the chiral
condensate $\pbp$.  Also the $U_A(1)$ symmetry, which is explicitely
broken by the axial anomaly, is expected to be effectively restored
at some temperature $T_A\geq T_c$.
Empirically also the Polyakov line has a rapid
increase at the same transition temperature as for chiral symmetry, 
indicating deconfinement:
it is not clear which relation exists between the chiral transition 
and the deconfining transition.
The transition line depicted in Fig.~\ref{PHDIA} is defined by
the maxima of a number of susceptibilities ($C_V$, $\chi_{m}$,
$\dots$), which indicate a rapid
variation of the corresponding parameters across the line:
the positions of these maxima all coincide within errors.

At $m \simeq 0$ it is possible to perform a renormalization group analysis plus
$\epsilon$-expansion techniques, assuming that the 
relevant degrees of
freedom for the chiral transition are scalar and pseudoscalar
fields~\cite{wilcz1,wilcz2,wilcz3}. The result is that if the
$U_A(1)$ symmetry is effectively restored, 
\ie if the $\eta '$ mass vanishes at $T_c$, then there is no IR stable
fixed point and the phase transition is first order; if this is not
the case an IR fixed point exists, which can produce a second order
phase transition in the $O(4)$ universality class.


If the first case is realized, the transition is first order also at 
$m \neq 0$ and most likely up to $m = \infty$.
In the second case a phase transition is present only at $m = 0$,
while a continuous crossover takes place as $m \neq 0$: that means that
one can move continuously from confined to deconfined and that 
no true order parameter exists. However this  
would be in contradiction with the experimental knowledge we
have about confinement: indeed the upper limit on the number
of free quarks per proton is $R = n_q/n_p \leq 10^{-27}$, while
$R \sim 10^{-12}$ is expected from the Standard Cosmological Model.  
A reduction factor $10^{-15}$ is difficult to explain in natural
ways unless $R = 0$, which means that confinement is 
related to some symmetry of the QCD vacuum and therefore 
the deconfining transition is associated to a change of symmetry
of the vacuum, \ie it is an order-disorder phase transition rather than
a continuous crossover.

The determination of the order of the chiral transition for $N_f = 2$
is therefore a fundamental issue, which can be settled by  
numerical simulations of lattice QCD.  
The problem has been investigated by several groups with
staggered~\cite{fuku1,fuku2,colombia,karsch1,karsch2,jlqcd,milc} and
Wilson~\cite{cp-pacs} fermions. The strategy used has been either to
look for signs of discontinuity at the transition, or to study the
scaling with respect to $m$ of different susceptibilities and of
the magnetic equation of state. No clear discontinuities
have been observed, but also no conclusive agreement of scaling with
$O(4)$ critical indexes.
A general tendency exists however in the community to consider the
chiral transition second order, and the line of Fig.~\ref{PHDIA} a
crossover.
We present the results of a big numerical effort using large
lattices, made in order to clarify the issue. 
A full account of our results can be found in~\cite{nostrolavoro}.
Like most of the other works
we use non improved Kogut--Susskind action, lattices $4 \times
L_s^3$ with $L_s = 12,16,20,24,32$, and the non exact R hybrid algorithm 
for $N_f = 2$~\cite{HybridR}. Scaling violations are expected
and a more careful study with $L_t = 6$, with an improved action and
algorithm is planned in order to control them.

\section{Results}

The theoretical tool to investigate the order of a phase transition is
finite size scaling. The extrapolation from
finite size $L_s$ to the thermodynamical limit is
governed by critical indexes, which identify the order and the
universality class of the transition.

Approaching the transition, for a higher order or weak first order
transition, the correlation length of the order parameter $\xi$ goes
large compared to the lattice spacing $a$, so that the dependence of
physical quantities on $a/\xi$ can be neglected. More precisely, if
${\cal L}/kT$ is the effective action (density of free energy)
\beq
\frac{\cal L}{kT} \simeq L_s^{-d} \phi \left(\frac{a}{\xi},  \frac{L_s}{\xi}, am_q L_s^{y_h} \right)
\label{scal1}
\eeq
the dependence on $a/\xi$ disappears as $T_c$ is
approached, since $\xi$ diverges as
\beq
\xi \simeq_{\tau \to \infty} \tau^{-\nu}
\eeq
where $\tau \equiv 1 - \frac{T}{T_c}$. 
Therefore around the chiral transition the system has two fundamental lengths:
the correlation length $\xi$ and the inverse quark mass
$1/m_q$. 
The variable $L_s/\xi$ can be traded
with $\tau L_s^{1/\nu}$ and the scaling law follows
\beq
\frac{\cal L}{kT} \simeq L_s^{-d} \phi \left(\tau L_s^{1/\nu}, am_q L_s^{y_h} \right) \, .
\label{scal2}
\eeq
As $\tau \to 0$ irrelevant terms can be neglected and the correlators
of the order parameter describe  the thermodynamics. The most
important quantity is however the specific heat, which shows the correct
critical behaviour independently of the identification of the order parameter.


The following scaling laws hold for the 
the specific heat and for the susceptibility of the order
parameter 
\beq
C_V - C_0 \simeq  L_s^{\alpha/\nu} \phi_c \left(\tau L_s^{1/\nu}, am_q L_s^{y_h} \right) \, ;
\label{scalcal} \\
\chi \simeq L_s^{\gamma/\nu} \phi_\chi \left(\tau L_s^{1/\nu}, am_q L_s^{y_h} \right) \, .
\label{scalord}
\eeq
$C_0$ stems from an additive renormalization. 
It is also possible to write them in the alternative form
\begin{eqnarray}
C_V - C_0 \simeq  L_s^{\alpha/\nu} \tilde\phi_c \left(\tau (am_q)^{-1/(\nu y_h)},am_q L_s^{y_h} \right)\label{scalcal2A} \\
\chi \simeq L_s^{\gamma/\nu} \tilde\phi_\chi \left(\tau (am_q)^{-1/(\nu y_h)}, am_q L_s^{y_h} \right) \, .
\label{scalord2A}
\end{eqnarray}

\begin{figure*}[t!]
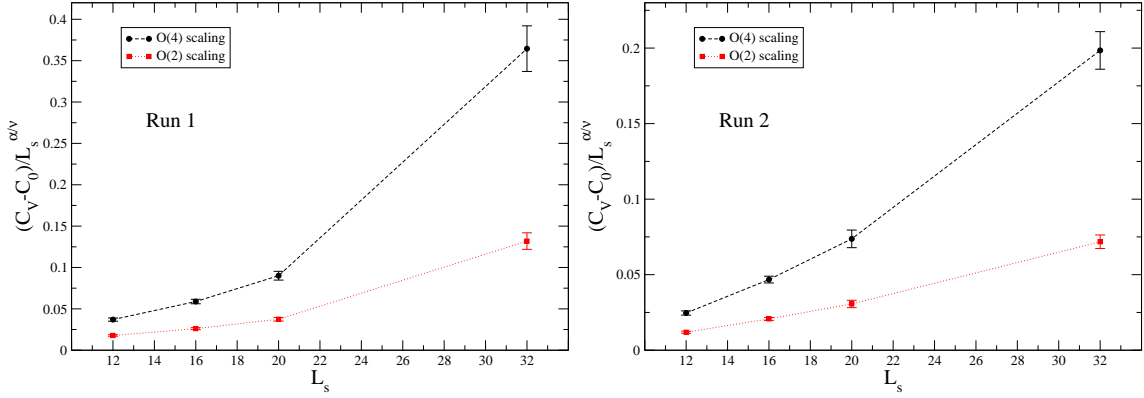

\includegraphics*[width=\columnwidth]{Cv_max_Run1.eps}
\includegraphics*[width=\columnwidth]{Cv_max_Run2.eps}\\
\vspace{-20pt}
\caption{Specific heat peak value for Run1
(left) and for Run2 (right), divided by the appropriate powers of
$L_s$ to give a constant. Both the $O(4)$
and $O(2)$ critical behaviors are displayed.}\label{R12}
\end{figure*}

\begin{table}[b!]
\caption{Critical exponents.}\label{CRITEXP}
\begin{tabular}{|c|c|c|c|c|c|}
\hline & $y_t$ & $y_h$ & $\nu$ & $\alpha$ & $\gamma$\\
\hline $O(4)$ & 1.34 & 2.49 & 0.75 & -0.23 & 1.48\\
\hline $O(2)$ & 1.49 & 2.49 & 0.67 & -0.01 & 1.33\\
\hline $MF$ & $3/2$ & $9/4$ & $2/3$ & 0 & 1\\
\hline $1^{st} Order$ & 3 & 3 & $1/3$ & 1 & 1\\
\hline
\end{tabular}
\end{table}


The phase transition is characterized by the critical
indexes: the values
relevant to our analysis are listed in
Table~\ref{CRITEXP}. $O(4)$ is the symmetry expected if the 
transition is second order, However lattice discretization 
can break it down to $O(2)$ for Kogut--Susskind fermions~\cite{karsch1} at
non zero lattice spacing.

Finite size scaling analysis is made difficult by the presence of two
independent scales. In the previous
literature the assumption has been usually made that the volume is
large enough so to neglect the dependence on $L_s$: 
since at fixed $am_q$, $\beta$
the susceptibilities must be analytic in the thermodynamical limit, 
as $L_s$ goes large the dependence on $am_q L_s^{y_h}$ must cancel the
dependence on $L_s$ in front of the scaling functions in
Eqs.~(\ref{scalcal}) and (\ref{scalord}). It follows that
\begin{eqnarray}
C_V - C_0 \simeq  (am_q)^{-\alpha/(\nu y_h)} f_c \left(\tau (am_q)^{-1/(\nu y_h)}\right) \label{scalcal1} \\
\chi \simeq  (am_q)^{-\gamma/(\nu y_h)} f_\chi \left(\tau (am_q)^{-1/(\nu y_h)}\right) \, .\label{scalord1}
\end{eqnarray}
The peak values of $(C_V - C_0)$ and of $\chi$ should then scale as
\beq
(C_V - C_0)_{\rm max} \propto (am_q)^{-\alpha/(\nu y_h)} \nonumber \\
\chi_{\rm max} \propto (am_q)^{-\gamma/(\nu y_h)}
\label{scalmax1}
\eeq
as $am_q\rightarrow 0$. The pseudocritical couplings, \ie
the positions of the maxima, should instead scale as 
\beq
\tau (am_q)^{-1/(\nu y_h)} = {\rm const} \label{pcscale} \, .
\eeq

\begin{figure*}[t!]
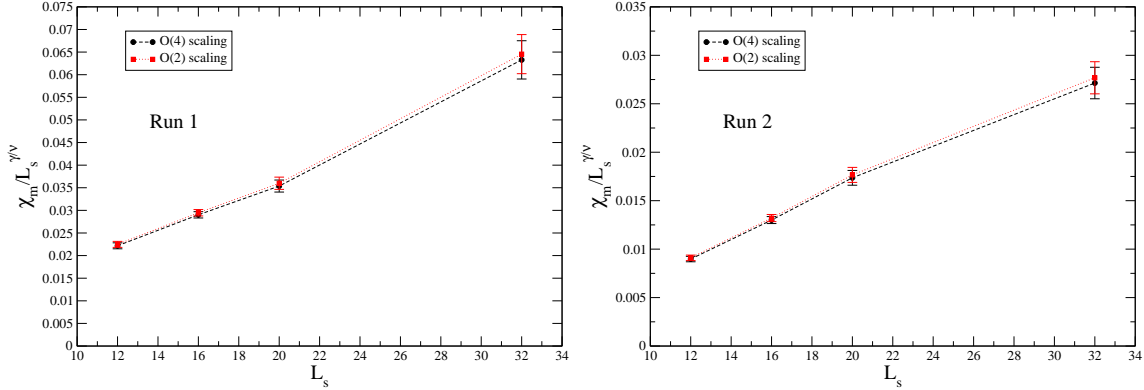

\includegraphics*[width=\columnwidth]{Chi_max_Run1.eps}
\includegraphics*[width=\columnwidth]{Chi_max_Run2.eps}
\vspace{-20pt}
\caption{The same as figure \ref{R12} for the chiral susceptibility
$\chi_m$.}\label{R12bis}
\end{figure*}

One can also consider to keep $\tau L_s^{1/\nu}$ fixed 
while taking
$a L_s \gg 1/m_\pi$. This assumption should work better if $L_s$ is
still comparable to the correlation length, which may be the case
close enough to the critical point. In this case the scaling laws are
\begin{eqnarray}
C_V - C_0 \simeq  (am_q)^{-\alpha/(\nu y_h)} f_c \left(\tau  L_s^{1/\nu} \right)
\label{scalcal2} \\
\chi \simeq  (am_q)^{-\gamma/(\nu y_h)} f_\chi \left(\tau  L_s^{1/\nu} \right) \, .
\label{scalord2}
\end{eqnarray}
Eqs.~(\ref{scalmax1}) stay unchanged, the positions
of the maxima scale as 
\beq
\tau L_s^{1/\nu} = {\rm const} \label{pcscale1} \, 
\eeq
while the widths of the peaks in this case are volume dependent.

We have instead followed a novel strategy which does not rely on any
assumption: in order to reduce the problem to one scale we have kept
fixed one of the scaling variables and we have studied the dependence
on the other.
As one can see in Table~\ref{CRITEXP}, the index $y_h$ is the
same for $O(4)$ and $O(2)$ symmetry. We have 
therefore made a number of simulations at
different values of $L_s$ and $am_q$ keeping $am_q L_s^{y_h}$ fixed
and assuming $y_h=2.49$ which corresponds to $O(4)$ or $O(2)$. In this
way as $L_s$ is increased, $am_q \to 0$, so that the infinite volume
limit corresponds to the chiral transition at $am_q = 0$.

From Eqs. (\ref{scalcal}) and (\ref{scalord}) it follows that the
maxima at constant $am_q L_s^{y_h}$ scale as
\beq
(C_V - C_0)_{\rm max} &\propto& L_s^{\alpha/\nu} \nonumber \\
\chi_{\rm max} &\propto& L_s^{\gamma/\nu} \, .\label{scalmax}
\eeq
as $L_s \rightarrow \infty$ and their positions scale as
\beq
\tau L_s^{1/\nu} = {\rm const}
\eeq
If $O(4)$ or $O(2)$ is the correct symmetry, the values of
$\alpha/\nu$ and $\gamma/\nu$ should be consistent with the
corresponding values listed in Table~\ref{CRITEXP}.

We have run two such sets of Monte
Carlo simulations, called in the
following Run1 and Run2, with $am_q L_s^{y_h}=74.7$
and $am_q L_s^{y_h}=149.4$ respectively. The spatial lattice sizes
$L_s$ used for each of the two sets are $L_s=12, 16, 20, 32$, the 
standard hybrid R algorithm~\cite{HybridR} has been used to update 
configurations.  

In Figs.~\ref{R12} and~\ref{R12bis} we show the peak values of the specific
heat and of the chiral susceptibility, divided by the power of 
$L_s$ appropriate for $O(4)$ or $O(2)$, as a function of $L_s$ (see
Eq.~\ref{scalmax}): scaling is clearly violated, $O(4)$ and
$O(2)$ universality classes are excluded by our data.

\begin{figure*}[t!]
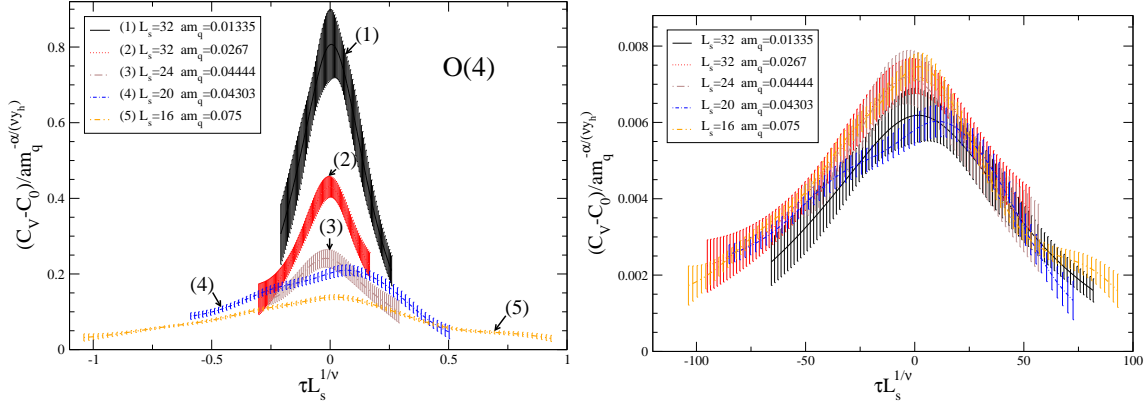

\includegraphics*[width=\columnwidth]{Cv_O4-Ls.eps} 
\includegraphics*[width=\columnwidth]{Cv_1st.eps}
\vspace{-20pt}
\caption{Comparison of specific heat scaling,  Eq.\ref{scalcal2}, for 
$O(4)$ (left) and first order (right).}\label{fullscal}
\end{figure*}

An alternative way to investigate the order of a phase transition is 
to study the scaling of pseudocritical couplings: one can try the two
alternative scaling laws of Eq.~(\ref{pcscale1}), 
\ie $\tau_c = k_\tau L_s^{1/\nu}$,  or of Eq.~(\ref{pcscale}), \ie
$\tau_c = k_\tau' (a m_q)^{1/(\nu y_h)}$ .
One has to be careful in defining the reduced temperature 
$\tau$ in presence of dynamical fermions, since in this case the 
physical temperature
$T = 1/ ( L_t a(\beta,m_q))$ is a function of both $\beta$ 
and $a m_q$, so that the reduced temperature $\tau$ can be expanded
as a power series in $(\beta - \beta_0)$ and in $a m_q$, where
$\beta_0$ is the chiral critical coupling. Only the
linear term in $\beta$ was considered in the previous literature. We
have found that the following terms are sufficient to fit the data  
\beq
\tau &\propto& (\beta_0 - \beta )+ k_m am_q + \nonumber \\ 
&+& k_{m^2} (am_q)^2 + k_{m\beta} am_q (\beta_0 - \beta ) \, .
\label{taudef2}
\eeq
As a result of our analysis we have found that it is not possible, 
within the present mass range, to
discriminate among the various possible critical behaviours by looking at
pseudocritical couplings only. We would like to remark that the
inclusion of other terms in Eq.~(\ref{taudef2}), besides the one 
linear in $\beta$ solely considered in previous literature, 
is crucial to obtain any scaling at all.

We have also tried to make a scaling analysis in the same
way as done in previous literature, 
\ie supposing that the lattice size is
much larger that all other relevant physical lengths.
No universality class is chosen a priori in this case and one can test all the 
possible critical behaviours. We have found that assuming that 
$a L_s \gg 1/m_\pi$ but still $L_s \sim \xi$, \ie 
Eqs.~(\ref{scalcal2}), (\ref{scalord2}) works better: this
is reasonable around the critical point, where $\xi$ goes large. 
We have added to the data
from Run1 and Run2, those from two other simulations performed at 
$L_s = 16$, $a m_q = 0.01335$ and $L_s = 24$, $a m_q = 0.04444$.
In Fig.\ref{fullscal} we show the scaling obtained for the specific
heat peak: $O(4)$ is again clearly excluded, while a good agreement is found 
with a weak first order critical behaviour. An analogous behaviour is
observed for the chiral susceptibility.

As a further test of scaling we check the equation of state
\beq
\pbp \simeq m^{1/\delta} f(\tau (a m_q)^{-1/(\nu y_h)})\label{eqstate}
\eeq
No scaling whatsoever is observed, neither
$O(4)$-$O(2)$ nor first order, if the raw measured data are introduced
in Eq.(\ref{eqstate}). After a proper subtraction~\cite{nostrolavoro} 
we have instead studied the following scaling law 
\beq
\pbp - \pbp_0 = (a m_q)^{1/\delta} F(\tau (a m_q)^{-1/\nu y_h})
\eeq
Results are 
shown in Fig.~\ref{eqstate}: again the first order behaviour describes
well the data while the second order is excluded.

\section{Conclusions}
We have argued that the study of the order of the chiral phase
transition for $N_f = 2$ is of fundamental importance to understand
confinement. 
By adopting a novel finite size scaling strategy which reduces
the analysis to a one scale problem, we have been able to 
show clear incompatibility of our data with $O(4)$ ($O(2)$) 
second order critical behaviour.
After introducing the correct definition of the reduced
temperature, we have shown that the analysis of pseudocritical 
coupling alone cannot discern between the possible critical
behaviours. 

\begin{figure*}[t!]
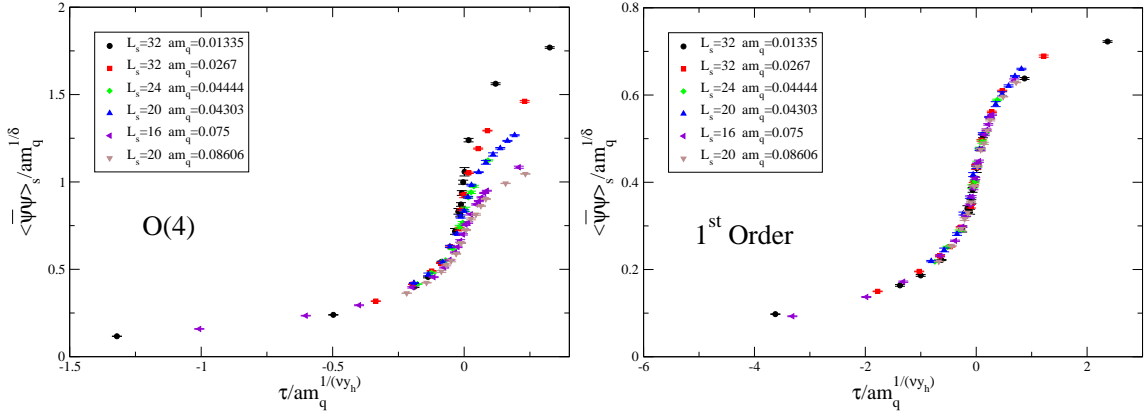

\includegraphics*[width=\columnwidth]{eqstO4.eps} 
\includegraphics*[width=\columnwidth]{eqst1st.eps}
\vspace{-20pt}
\caption{Scaling of the equation of state for 
$O(4)$ (left) and first order (right).}\label{eqstate}
\end{figure*}

We have also repeated the same finite size scaling analysis
performed in previous literature, \ie assuming that
the thermodynamical limit has been reached, and again
we have found disagreement with $O(4)$ ($O(2)$) second order critical 
behaviour and consistency with a weak first order critical
behaviour both in the scaling of susceptibilities and in the equation
of state. This would be in agreement with confinement being an 
absolute property of matter related to some symmetry and with the
deconfining transition being order-disorder, even if we have  
still not found any clear evidence for discontinuities in physical
observables. 

We would like to remark that our simulations, like in most of previous
works, have been performed with
the standard staggered action at $L_t = 4$ and with the non exact
Hybrid R algorithm: scaling corrections could be important in this
case and  the issue is surely still open: we plan in the future 
to investigate it more deeply by making simulations with improved actions and
algorithms and with $a m_q L_s^{y_h}$ fixed according to 
first order.

\section*{Acknowledgements}
This work has been partially supported by MIUR, Program ``Frontier
problems in the Theory of Fundamental Interactions''.

\end{document}